%% file: 0_template_sigchi.tex
\documentclass[sigconf]{acmart}
\AtBeginDocument{%
  \providecommand\BibTeX{{%
    \normalfont B\kern-0.5em{\scshape i\kern-0.25em b}\kern-0.8em\TeX}}}
\usepackage{blindtext}
\makeatletter
\newcommand{\global@insert}[2]% #1=box number, #2=vertical list
{\bgroup
  \setbox\@tempboxa=\box#1
  \global\setbox#1=\vbox{\unvbox\@tempboxa #2}
\egroup}

\long\def\@footnotetext#1{\global@insert\footins{%
 \reset@font\footnotesize
 \interlinepenalty\interfootnotelinepenalty
 \splittopskip\footnotesep
 \splitmaxdepth \dp\strutbox \floatingpenalty \@MM
 \hsize\columnwidth \@parboxrestore
 \protected@edef\@currentlabel{%
 \csname p@footnote\endcsname\@thefnmark
 }%
 \color@begingroup
 \@makefntext{%
 \rule\z@\footnotesep\ignorespaces#1\@finalstrut\strutbox}%
 \color@endgroup}}%
\makeatother
\newcommand{\boldpara}[1]{\medskip\noindent\textbf{{#1}}}

\usepackage{graphicx}
\usepackage{caption}
\usepackage{subcaption}
\usepackage{enumitem}

\setcopyright{acmcopyright}
\copyrightyear{2018}
\acmYear{2018}
\acmDOI{10.1145/1122445.1122456}

\acmConference[]{}{}{}
\acmBooktitle{}
\acmPrice{}
\acmISBN{}
\setcopyright{none}
\renewcommand\footnotetextcopyrightpermission[1]{}
\begin{document}

\title[Viewership-centric Automated Content Curation]{
Scaling New Peaks:\\
A Viewership-centric Approach to Automated Content Curation}


% \author{AUTHOR INFORMATION REDACTED}
\author{Subhabrata Majumdar}
\email{subho@att.com}
\affiliation{%
  \institution{Data Science and AI Research, AT\&T}
  \country{}
}
\author{Deirdre Paul}
\authornote{contributed to work while at AT\&T Labs -- Research.}
\email{dedepaul@gmail.com}
\affiliation{
\country{}
}
\author{Eric Zavesky}
\email{ezavesky@att.com}
\affiliation{%
  \institution{Data Science and AI Research, AT\&T}
  \country{}
}

% \renewcommand{\shortauthors}{Redacted}
\renewcommand{\shortauthors}{Majumdar et al}

\begin{abstract}
Summarizing video content is important for video streaming services to engage the user in a limited time span. To this end, current methods involve manual curation or using passive interest cues to annotate potential high-interest segments to form the basis of summarized videos, and are costly and unreliable. We propose a viewership-driven, automated method that accommodates a range of segment identification goals. Using satellite television viewership data as a source of ground truth for viewer interest, we apply statistical anomaly detection on a timeline of viewership metrics to identify `seed’ segments of high viewer interest. These segments are post-processed using empirical rules and several sources of content metadata, e.g. shot boundaries, adding in personalization aspects to produce the final highlights video.

To demonstrate the flexibility of our approach, we present two case studies, on the United States Democratic Presidential Debate on 19th December 2019, and Wimbledon Women’s Final 2019. We perform qualitative comparisons with their publicly available highlights, as well as early vs. late viewership comparisons for insights into possible media and social influence on viewing behavior.
\end{abstract}





\maketitle

\input{1_intro}

\input{2_system}

\input{3_results}

\input{4_conc}

\bibliographystyle{ACM-Reference-Format}
\bibliography{0_template_sigchi}

\end{document}

%% file: 1_intro.tex
\section{Introduction}\label{sec:intro}

With the democratization of video content creation, increase in on-demand video consumption, and proliferation of video content forms and channels, the need to drive user engagement by providing content that is succinct, interesting and personalized is expected to grow. To this end, methods to differentiate, personalize and contextualize a provider’s {\it summarized} video content in order to engage the user in a limited time span are increasingly critical.

Current methods for summarizing video content or identifying content highlights include manual extraction, analysis of video components and metadata such as images, facial expressions or crowd noise to infer emotion or excitement \cite{merler_automatic_2019} and analysis of secondary data such as social media \cite{gotsentiment,wongsuphasawat_how_2016}. These methods can be costly (with respect to resource demands or response latency), and their utility is severely limited as the fundamental inputs are not available for the majority of consumed content. Alternate methods have also been developed for generating summaries or trailers of theatrical content and are an active topic in computer vision---where solutions are almost entirely automated \cite{yale_song_tvsum_2015, otani_rethinking_2019, park_adversarial_2019, baraldi_recognizing_2016, smith_ibm_2016}. While continuously improving, these methods often lack the semantic of emotional connections to the content that often determine true user engagement. Even though some content creators may be able to a priori annotate potential regions of high or low interest that form the basis of summarized videos, the volume and variety of content demands an automated and scalable approach.

\begin{figure*}[htbp]
\centering
\begin{subfigure}[b]{.49\linewidth}
\caption{Segment viewership metrics}
\includegraphics[width=\linewidth, clip, trim=200pt 105pt 190pt 105pt ]{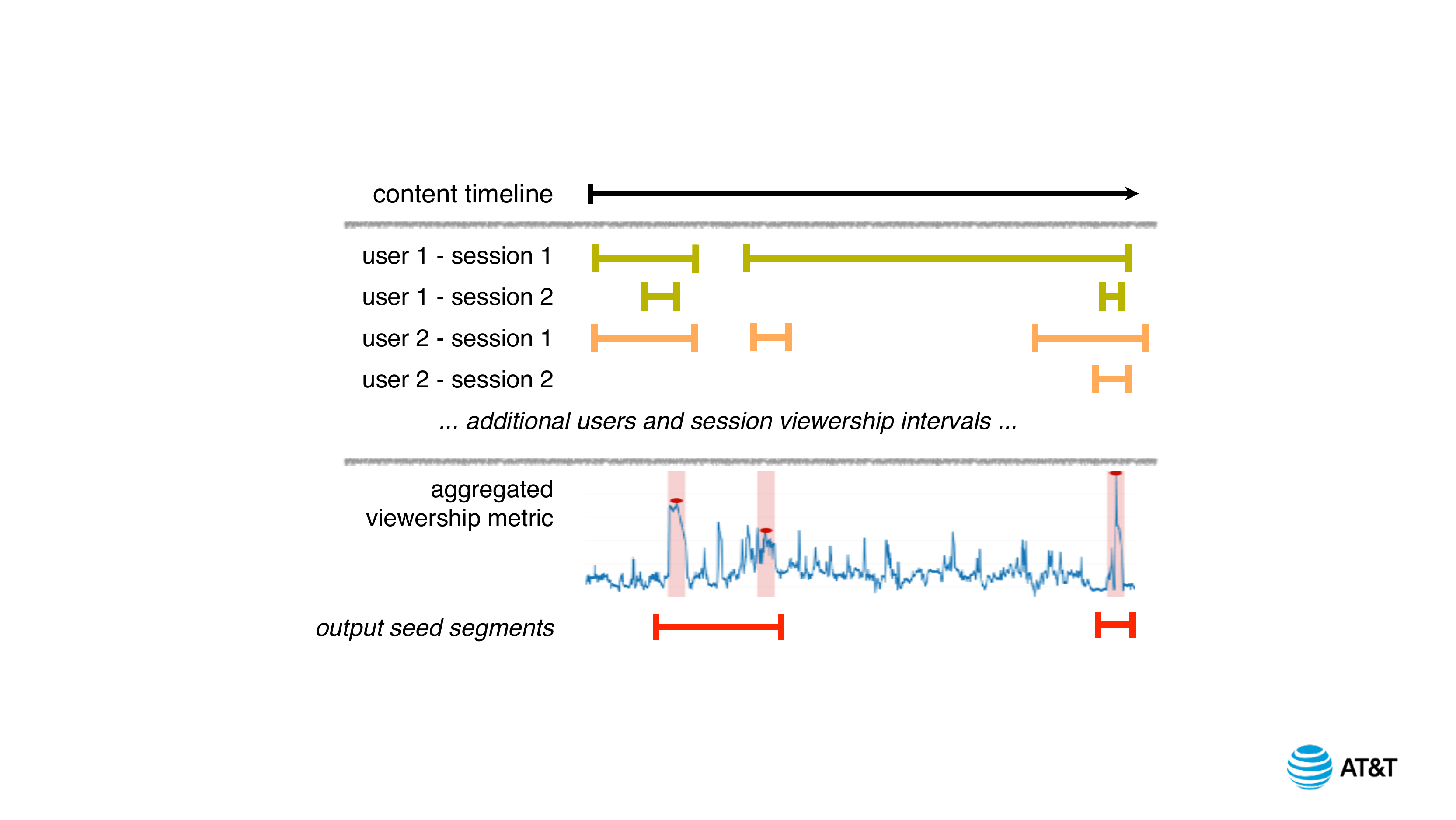}
\label{fig:seg-viewer}
\end{subfigure}
\begin{subfigure}[b]{.49\linewidth}
\caption{Segment metadata}
\includegraphics[width=\linewidth, clip, trim=200pt 105pt 190pt 105pt ]{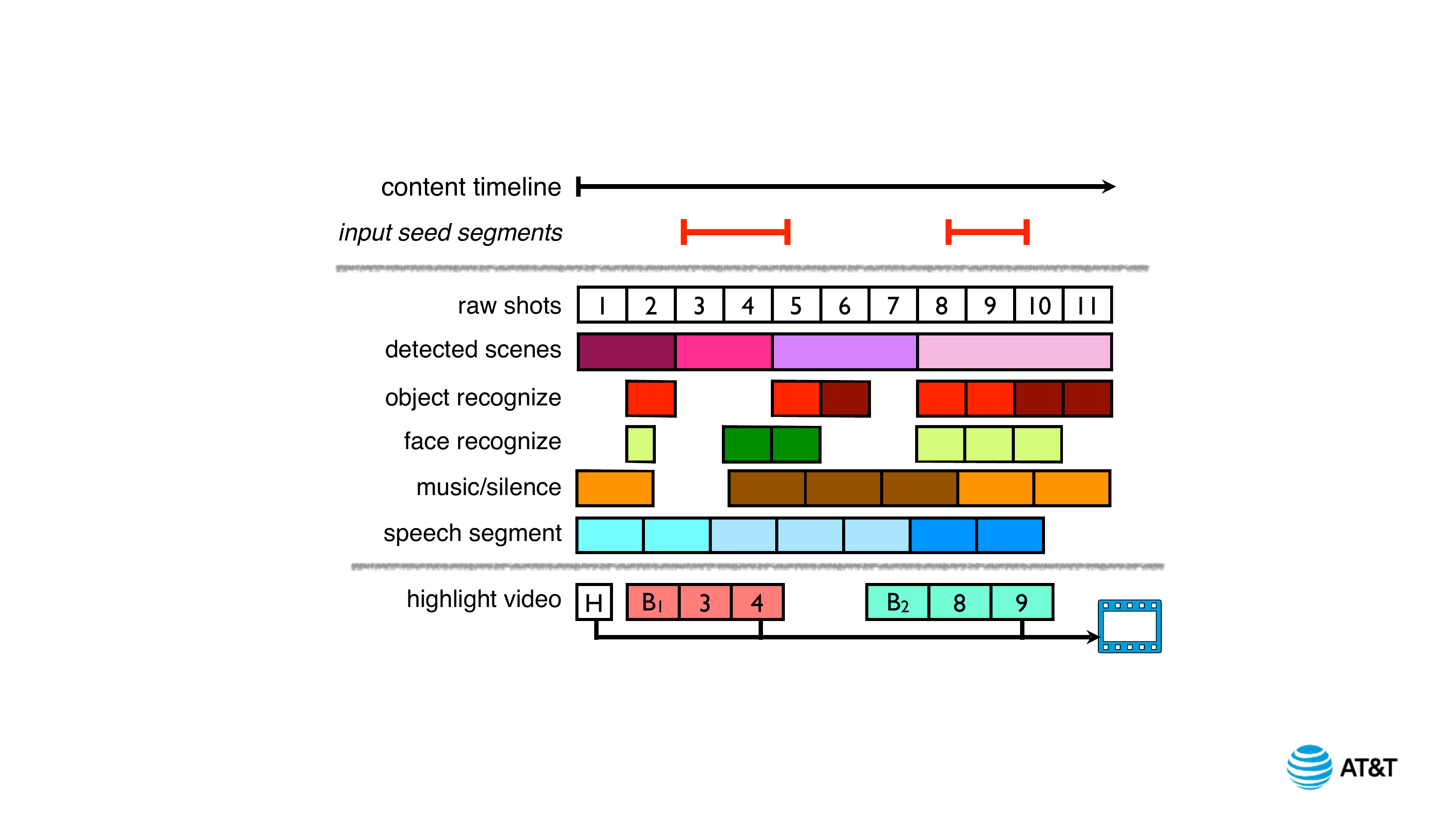}
\label{fig:seg-metadata}
\end{subfigure}
\caption{Schematic of viewership metric derivation and metadata alignment. In panel~\ref{fig:seg-viewer}, peak segments in the viewership timeline are aggregated and detected. Using this time-based output, selected aligned shots (3,4) and (8, 9, 10) cover input seeds from the timeline in panel~\ref{fig:seg-metadata}. Segments are stitched together and optionally interspersed with header ($H$) and bumpers ($B_1,B_2$) in a highlights.}
\label{fig:integ}
\end{figure*}

We propose a viewership-driven, automated method that accommodates a range of segment identification goals as depicted in Fig.~\ref{fig:integ}. While we mainly focus on content summarization in this work, applications of this are numerous and currently under-served by state-of-the-art methods: personalized curation, targeted advertising, higher quality new content, alternate content and product placement optimization. Specifically, this method presents powerful advantages over prior summarization techniques.

\noindent{\bf Reduction of domain expertise burden.} As the diversity of available content grows (how-to, theatrical, personal video logs, reality-filming), segment generation from viewership behaviors is more likely to engage the average end-user. 
% , generating richer content products based on viewrship . With more vertical integration in media companies, we can build on this curated video foundation and create richer products, leveraging data from new sources. 

\noindent{\bf Detection of segments without bias and minimal observations.} This method works for sparsely viewed content as well as popular content, and is free of the bias of a manual curator.\\
% (c)	different versions of the high/low interest events can be generated by 
\noindent{\bf Many personalization opportunities.} Tuning the anomaly algorithm for preferences and situational contexts (user interest, short or long duration, time, location, consumption medium) personalizes each generated event.
% personalization and contextualization is baked into the process.

Although this paper presents a method with early findings, we seek to answer these research questions through an evaluation of automated- and human-generated content highlight videos.

\begin{itemize}[leftmargin=*]
\item Can a highlight generation algorithm accommodate events that may be temporally uncorrelated (e.g. live moments in content have no dependence on each-other) or highly structured/evolving (e.g. scoring moments in sports or critical plot points in a drama)?

\item Can an automated highlight generation algorithm achieve segment-level parity with content from human curator driven sources?
\end{itemize}

%% file: 2_system.tex
\section{Technology and System Design}\label{sec:sys}
Our method uses anonymized and aggregated viewership data from a leading satellite broadcast provider as a source of ground truth for user interest. We hypothesize that repeatedly watching a content segment indicates the viewer is highly interested in that content.  

\subsection{Algorithms}\label{sec:algorithms}
Starting with a novel viewership metric derived from repeat watching behavior, metadata produced by content analysis tools operating on the video itself are used to segment highlights, their relevance, and create excerpt videos.  

% Modern systems built for summarization and curation of content heavily rely on human exemplars for training.  In the space of summary videos (trailers, stories, or brief recap), the automated solutions rely on heuristics about object uniqueness, actor appearances \cite{yale_song_tvsum_2015}, \cite{otani_rethinking_2019} or require explicit examples to learn   \cite{park_adversarial_2019}. This work focuses on the space of curated, snackable (or bite-sized) content segments that typically range from 30 to 90 seconds.  The proposed system
% %,\popcorn, 
% extends works above, utilizing them at a more elemental level, as illustrated in figure \ref{fig:popcorn_processing}.  Here, a number of automatic classifiers based on visual content produce semantic tags (e.g. \textit{smiling, climbing, explosion, underwater}) with specific time-code references to the content, generally called "timed metadata".  This work uses timed metadata as a tool to generate a variety of \textbf{cut} at times in the content and defers discussions cut diversity and model training details to other works.

\boldpara{Metadata Tags.}
We associate the content timeline with metadata that is either externally supplied or automatically generated, such as actors, scenes, emotions, and demographic preferences.  Previous statistical methods utilized actor and object count to infer importance to a scene \cite{yale_song_tvsum_2015, otani_rethinking_2019}, but this requires domain-specific labels (e.g. sports, drama, etc.) to guarantee metadata importance.  Similar to these works, metadata from a mix of internally created (emotion, theme, specific actor identities) and cloud vendor computed \cite{azure,goog,aws} are pooled for each video with exemplars depicted in Fig.~\ref{fig:seg-metadata}.  The flexibility of this integration strategy better accommodates future personalization and contextualization after learning relevant metadata tags for a specific user.

\boldpara{Viewership Metric.}
For a specific video asset, we start from a set of relevant viewers, for example,  viewers that watched a certain minimum duration,in a specific geographical area, with specific viewing history, or those watching within a certain time of show airing. This critical viewer base allows independence from metadata-only methods that generate implied interest regions \cite{merler_automatic_2019}, or metadata with no clear connection to user interest \cite{gotsentiment,wongsuphasawat_how_2016}. 

Our viewership metric is the percentage of viewer base that repeatedly watched a content segment of pre-specified length over Digital video recorder (DVR). By replicating this process for all possible segments across the timeline of the television show, we generate a time series of the metric. We then apply the IQR method of statistical anomaly detection \cite{anombook} to identify ‘seed’ segments of peak user interest, corresponding to time points belonging to higher percentiles of the normalized time series. We postprocess the seed segments with the help of content metadata like shot boundaries and empirical rules like joining closely situated segments or expanding very short segments to a default length of 15 seconds to produce coherent time segments in summarized highlights (Fig.~\ref{fig:seg-viewer}).

\subsection{Content Type Invariance}
%Since AT&T owns both the content creation (WarnerMedia) and broadcast (DIRECTV, HBO Max) platforms, it is uniquely positioned to benefit from this cutting edge technology.
The system leverages the viewership metric and metadata alignment defined in the previous section to detect interesting events inside content, which we know to be of high interest.  We then select segments corresponding to a tag (or combination of tags) relevant to a specific user or user segment to curate into the highlights alongside the ‘crowd-sourced’ high-interest segments.  

Targeting a core research question, we designed the algorithm to accommodate content that contained moments that are temporally uncorrelated (reality shows, news reports, a documentary) and those that are highlight correlated (sporting events, plots points in a drama, etc.).  Some approaches derive complex modeling within a core algorithm to accommodate this variance \cite{baraldi_recognizing_2016, park_adversarial_2019}, but we chose to modulate input preprocessing and thereby maintain complexity and fidelity of the underlying algorithm based on the viewership metric.  While this choice does make an assumption that new content will have identifiers that aid automated determination of its uncorrelated or correlated nature, this burden is small and knowledge of content source alone (professional media production, user-generated content channels, social media) may be already sufficient.
% As an early work, we present a method that has distinct advantages over existing practices for the generation of content highlights from different types of content. 
We present and evaluate two variants of our methodology to demonstrate this event-level flexibility. 
% Based on different personalization and metadata integration scenarios, we designed two variants of our curation process. 
% These variants have two major common components -
The variants utilize a common aggregated timeline of a viewership metric and metadata tags, but these components work together in different fashion to generate multiple highlights videos of the same content.

\begin{figure*}[t]
\centering
\begin{subfigure}[b]{.99\linewidth}
\caption{Variant 1 (V1, content) pipeline}
\includegraphics[width=\linewidth,clip, trim=20pt 160pt 20pt 160pt  ]{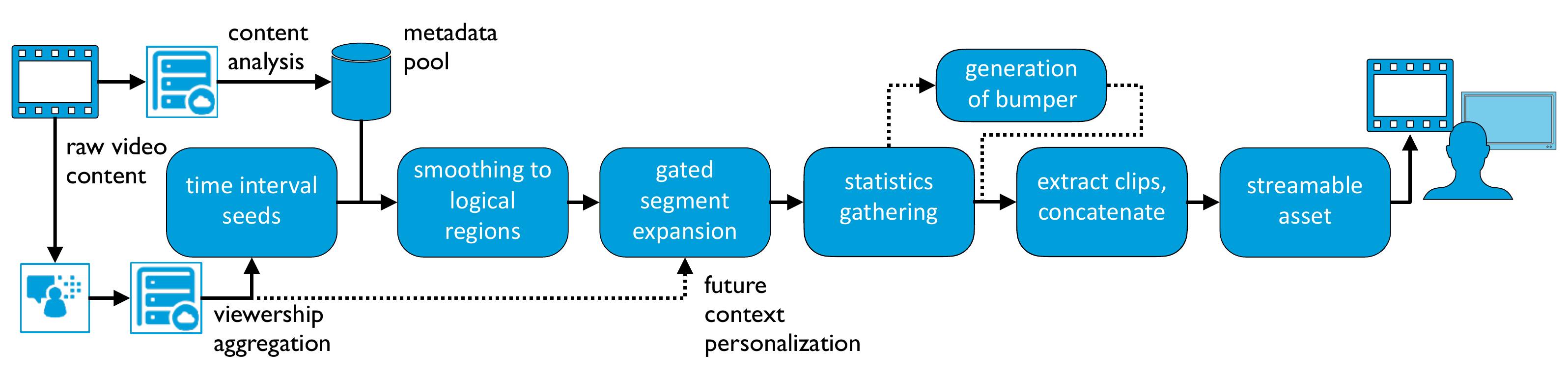}
\label{fig:seg-variant1}
\end{subfigure}
\begin{subfigure}[b]{.99\linewidth}
\caption{Variant 2 (V2, events) pipeline}
\includegraphics[width=\linewidth,clip, trim=20pt 130pt 20pt 130pt  ]{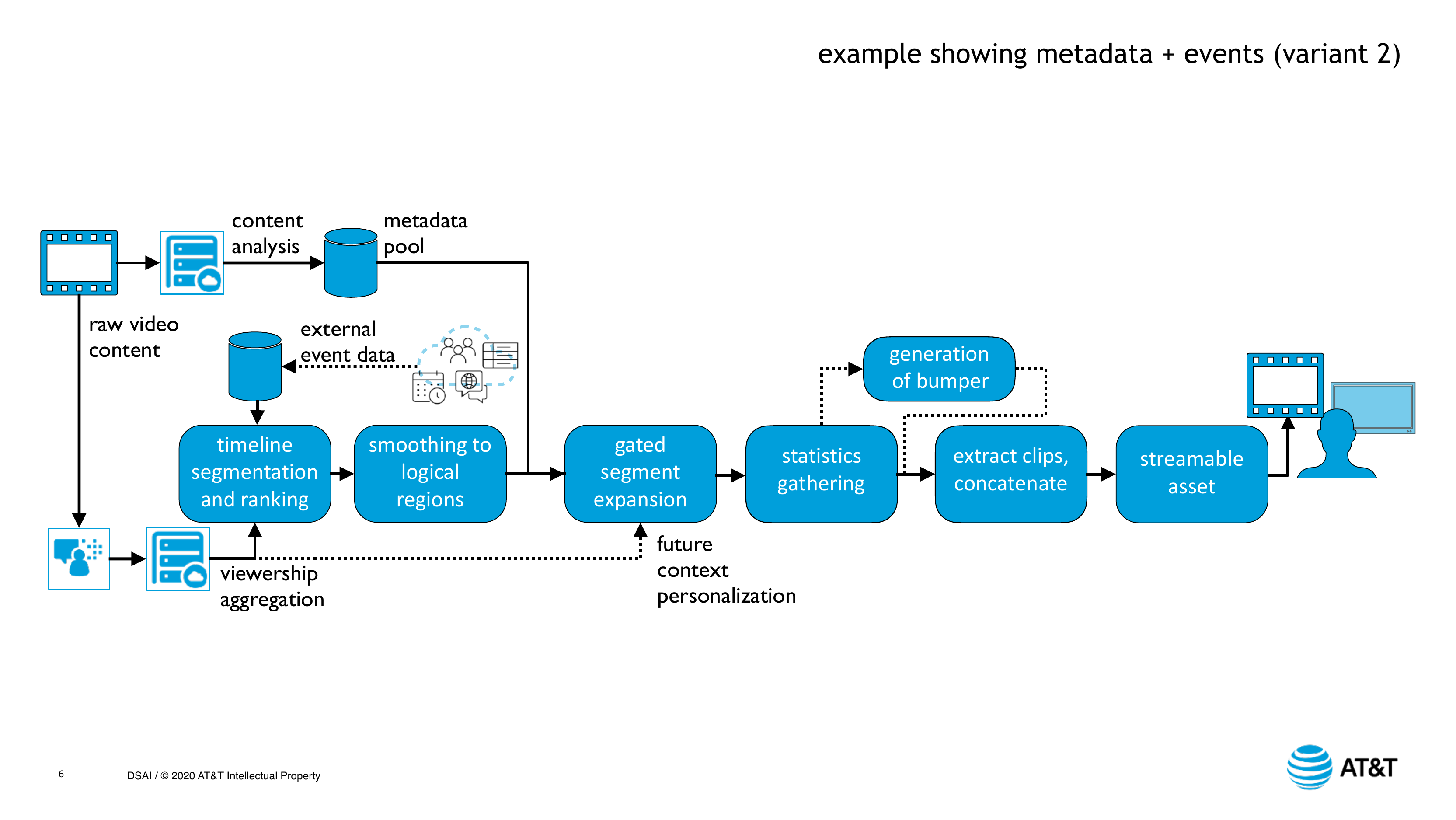}
\label{fig:seg-variant2}
\end{subfigure}
\caption{Viewership-centric highlights generation pipelines. V1 uses anomaly detection on viewership metric(s) and metadata integration. V2 uses external event data as a preprocessing step to segment and rank content.}
\label{fig:seg-algorithms}
\end{figure*}

\begin{table*}[t]
\centering
\scalebox{.9}{
\begin{tabular}{cl|cccccc}
\hline
%-----
\parbox[t]{1cm}{\centering{\bf Key}} &
\parbox[t]{5cm}{{\bf Description}} &
\parbox[t]{0.8cm}{\centering{\bf V1}} &
\parbox[t]{1.3cm}{\centering{\bf Nymag}} &
\parbox[t]{1.1cm}{\centering{\bf CNN}} &
\parbox[t]{1cm}{\centering{\bf Time}} &
\parbox[t]{1.6cm}{\centering{\bf Guardian}} &
\parbox[t]{1cm}{\centering{\bf NYT}}
\\\hline \hline
%-----
\parbox[t]{.5cm}{\centering 1} &
\parbox[t]{4cm}{Candidates on impeachment} &
\parbox[t]{1cm}{\centering Y} &
\parbox[t]{1cm}{\centering Y} &
\parbox[t]{1cm}{} &
\parbox[t]{1cm}{\centering Y} &
\parbox[t]{1cm}{\centering Y} &
\parbox[t]{1cm}{}\\\hline
%-----
\parbox[t]{.5cm}{\centering 2} &
\parbox[t]{4cm}{Candidates on economy} &
\parbox[t]{1cm}{\centering Y} &
\parbox[t]{1cm}{} &
\parbox[t]{1cm}{} &
\parbox[t]{1cm}{} &
\parbox[t]{1cm}{} &
\parbox[t]{1cm}{}\\\hline
%-----
\parbox[t]{.5cm}{\centering 3} &
\parbox[t]{8cm}{Warren silences tax plan critics:
``Oh, they're just wrong!"} &
\parbox[t]{1cm}{\centering Y} &
\parbox[t]{1cm}{\centering Y} &
\parbox[t]{1cm}{} &
\parbox[t]{1cm}{} &
\parbox[t]{1cm}{} &
\parbox[t]{1cm}{}\\\hline
%-----
\parbox[t]{.5cm}{\centering 4} &
\parbox[t]{5cm}{Candidates on climate change} &
\parbox[t]{1cm}{\centering Y} &
\parbox[t]{1cm}{} &
\parbox[t]{1cm}{\centering Y} &
\parbox[t]{1cm}{\centering Y} &
\parbox[t]{1cm}{} &
\parbox[t]{1cm}{}\\\hline
%-----
\parbox[t]{.5cm}{\centering 5} &
\parbox[t]{8cm}{Yang, Sanders on issues related to People of Color} &
\parbox[t]{1cm}{\centering Y} &
\parbox[t]{1cm}{} &
\parbox[t]{1cm}{} &
\parbox[t]{1cm}{\centering Y} &
\parbox[t]{1cm}{} &
\parbox[t]{1cm}{}\\\hline
%-----
\parbox[t]{.5cm}{\centering 6} &
\parbox[t]{8cm}{Sanders answers age question: ``And I'm white as well!"} &
\parbox[t]{1cm}{\centering Y} &
\parbox[t]{1cm}{\centering Y} &
\parbox[t]{1cm}{\centering Y} &
\parbox[t]{1cm}{} &
\parbox[t]{1cm}{\centering Y} &
\parbox[t]{1cm}{\centering Y}\\\hline
%-----
\parbox[t]{.5cm}{\centering 7} &
\parbox[t]{4cm}{Biden says ``I was joking"} &
\parbox[t]{1cm}{\centering Y} &
\parbox[t]{1cm}{} &
\parbox[t]{1cm}{\centering Y} &
\parbox[t]{1cm}{} &
\parbox[t]{1cm}{\centering Y} &
\parbox[t]{1cm}{}\\\hline
%-----
\parbox[t]{.5cm}{\centering 8} &
\parbox[t]{9cm}{Warren on age:``I'd also be the youngest woman ever inaugurated"
} &
\parbox[t]{1cm}{\centering Y} &
\parbox[t]{1cm}{\centering Y} &
\parbox[t]{1cm}{} &
\parbox[t]{1cm}{\centering Y} &
\parbox[t]{1cm}{\centering Y} &
\parbox[t]{1cm}{\centering Y}\\\hline
%-----
\parbox[t]{.5cm}{\centering 9} &
\parbox[t]{4cm}{Warren on billionaire donors} &
\parbox[t]{1cm}{\centering Y} &
\parbox[t]{1cm}{} &
\parbox[t]{1cm}{} &
\parbox[t]{1cm}{} &
\parbox[t]{1cm}{} &
\parbox[t]{1cm}{}\\\hline
%-----
\parbox[t]{.5cm}{\centering 10} &
\parbox[t]{4cm}{Warren and Buttigieg argue} &
\parbox[t]{1cm}{\centering Y} &
\parbox[t]{1cm}{\centering Y} &
\parbox[t]{1cm}{\centering Y} &
\parbox[t]{1cm}{\centering Y} &
\parbox[t]{1cm}{\centering Y} &
\parbox[t]{1cm}{\centering Y}\\\hline
%-----
\parbox[t]{.5cm}{\centering 11} &
\parbox[t]{5cm}{Sanders on billionaire donors"} &
\parbox[t]{1cm}{\centering Y} &
\parbox[t]{1cm}{} &
\parbox[t]{1cm}{} &
\parbox[t]{1cm}{} &
\parbox[t]{1cm}{} &
\parbox[t]{1cm}{\centering Y}\\\hline
%-----
\parbox[t]{.5cm}{\centering 12} &
\parbox[t]{8cm}{Yang on excluding women: ``we kind of become morons"} &
\parbox[t]{1cm}{\centering Y} &
\parbox[t]{1cm}{} &
\parbox[t]{1cm}{} &
\parbox[t]{1cm}{} &
\parbox[t]{1cm}{} &
\parbox[t]{1cm}{}\\\hline
%-----
\parbox[t]{.5cm}{\centering 13} &
\parbox[t]{5cm}{Klobuchar and Buttigieg argue} &
\parbox[t]{1cm}{\centering Y} &
\parbox[t]{1cm}{\centering Y} &
\parbox[t]{1cm}{\centering Y} &
\parbox[t]{1cm}{\centering Y} &
\parbox[t]{1cm}{\centering Y} &
\parbox[t]{1cm}{\centering Y}\\\hline
%-----
\parbox[t]{.5cm}{\centering 14} &
\parbox[t]{4cm}{Biden on Afghanistan} &
\parbox[t]{1cm}{\centering Y} &
\parbox[t]{1cm}{} &
\parbox[t]{1cm}{} &
\parbox[t]{1cm}{} &
\parbox[t]{1cm}{} &
\parbox[t]{1cm}{}\\\hline
%-----
\parbox[t]{.5cm}{\centering 15} &
\parbox[t]{4cm}{Candidates on China} &
\parbox[t]{1cm}{} &
\parbox[t]{1cm}{} &
\parbox[t]{1cm}{} &
\parbox[t]{1cm}{\centering Y} &
\parbox[t]{1cm}{} &
\parbox[t]{1cm}{}\\\hline
%-----
\parbox[t]{.5cm}{\centering 16} &
\parbox[t]{4cm}{Biden stuttering} &
\parbox[t]{1cm}{} &
\parbox[t]{1cm}{\centering Y} &
\parbox[t]{1cm}{} &
\parbox[t]{1cm}{} &
\parbox[t]{1cm}{\centering Y} &
\parbox[t]{1cm}{}\\\hline
%-----
\parbox[t]{.5cm}{\centering 17} &
\parbox[t]{8cm}{Biden and Sanders argue on healthcare} &
\parbox[t]{1cm}{\centering Y} &
\parbox[t]{1cm}{\centering Y} &
\parbox[t]{1cm}{\centering Y} &
\parbox[t]{1cm}{} &
\parbox[t]{1cm}{} &
\parbox[t]{1cm}{\centering Y}\\\hline
%-----
\parbox[t]{.5cm}{\centering 18} &
\parbox[t]{8cm}{Yang wants to give other candidates his book} &
\parbox[t]{1cm}{\centering Y} &
\parbox[t]{1cm}{} &
\parbox[t]{1cm}{\centering Y} &
\parbox[t]{1cm}{} &
\parbox[t]{1cm}{} &
\parbox[t]{1cm}{}\\\hline
%-----
\parbox[t]{.5cm}{\centering 19} &
\parbox[t]{4cm}{Closing statements} &
\parbox[t]{1cm}{\centering Y} &
\parbox[t]{1cm}{\centering Y} &
\parbox[t]{1cm}{\centering Y} &
\parbox[t]{1cm}{} &
\parbox[t]{1cm}{} &
\parbox[t]{1cm}{}\\
\hline
\end{tabular}
}
    \caption{Chronologically ordered clips across different highlight videos/summaries of the 19th Dec. 2019 US Democratic Debate.}
    \label{tab:debatecomp}
\end{table*}

\boldpara{Variant 1 (V1, content).}
In this variant (Fig. \ref{fig:seg-variant1}), the viewership metric drives time interval detection, followed by smoothing to detected logical shot boundaries (i.e. expanding/shortening the segment to snap to closest shot start/end).  The additional content metadata provides a rich selection of alternate features that can be used to fine-tune the start and end of a segment boundary (e.g. getting more context for the highlight, focusing on a face if detected in the scene). Work in this area utilizing content metadata as sole grounds for detecting time intervals also exists, but most systems require supervised examples from the user in pairwise assertions \cite{baraldi_recognizing_2016} or fully prepared video exemplars \cite{wang_learning_2020}. Specifically exploring clip personalization with this data, a related prior work by the authors conducted user studies for incorporating additional contextual segments \cite{snackable}.  We consider V1 as a content-invariant baseline in our case studies (Section~\ref{sec:res}) and utilize to generate highlight videos by first joining/expanding seed segments, then smoothing to shot boundaries from content metadata.

\boldpara{Variant 2 (V2, events).}
To incorporate external event metadata, the V1 pipeline is slightly modified to first segment the viewership metric timeline, then pick the segments with highest scores (Fig. \ref{fig:seg-variant2}). Such content segmentation may either be available publicly or can be obtained from partnering with content creators. Examples of useful segmenting include play-by-play data for baseball \cite{retro} and American football \cite{nflsavant}, and point-level analysis by IBM SlamTracker for Tennis \cite{slam}. Essentially, for content events that permit a natural segmentation, such as points or plays in a sports game, instead of picking high-interest parts in the timeline of rewatch scores then post-processing them (V1), a second curation approach is to score each pre-defined segment and pick highest scoring segments. Similar to V1, optional personalization of events can be achieved by incorporating additional high-scoring contextual events---such as most rewatched points that a specific player won, or a collection of most rewatched third down pickups by a quarterback across multiple games of American football.

%% file: 3_results.tex
\section{Case studies}\label{sec:res}

We now present two case studies for granular comparison of auto-generated highlights videos by our system with publicly available highlights of the same content. To emphasize the general nature of V1, in both case studies we use videos auto-curated using this method (Figure~\ref{fig:seg-variant1}). For the second application we qualitatively compare V1 and V2 curated videos.

% \subsection{US Democratic Presidential Debate}\label{sec:debates}
\boldpara{US Democratic Presidential Debate.}
We picked the United States Democratic Debate on 19th December 2019 as a case study into the `hands-free' segment detection and personalization aspects of V1. We selected this media because it was one of the most widely viewed events during Dec. 2019, per aggregated viewership data. It also demonstrates lower temporal dependence in the content (e.g. debate point 11 in Table~\ref{tab:debatecomp} isn't causal for 16) and we also observe heavy skipping/rewatching behavior.
% \boldpara{Design}
To navigate potential curator bias, we chose summary articles and videos from diverse sources \cite{cnn,nyt,nymag,time,guardian}: a news outlet (CNN), print media (New York Times, New York Magazine), and international outlets (Time, Guardian). %\cite{cnn,nyt,nymag,time,guardian}
%\block{
%talk about how we picked this specific video and found the websites for comparison; specifically (because this event has some political / curator potential bias) how did we avoid matching to a "democrat favoring" algorithm; did we have to exclude certain topics, how did you make sure they covered the entire event; }

% \boldpara{Methods}
To begin the comparison process, we went through the V1 highlights video and annotated each clip with short descriptions, then similarly partitioned each external summary. For text summaries (Nymag, CNN, Time), these partitions corresponded to subject or event-specific sections of each article, while partitions of video summaries (Guardian, NYT) comprised of component clips of each video. The intersection of a V1 clip with a text summary partition corresponded to all tokenized and lemmatized words in a clip annotation being present in the tokenized and lemmatized text under that partition of the article. The intersection of a V1 clip with a video summary clip corresponded to an overlap of at least 10 seconds between auto-generated closed-captions of the two clips from cloud vendors mentioned in Section~\ref{sec:algorithms}.

% \block{I did this `manually', but assuming this is plausible? open to other ways of finessing it. also what is a method for auto-generating captions?}
%\block{how did we compare the segments to eachother, was there multiple people annotating the different segments? was there potential bias in selecting the segments or did you use very similar image matching? how what was the method used to guarantee coverage of the overlapping events}

% \boldpara{Results}
Among the 17 clips in V1, 13 (76\%) were present in at least one other summary. On the other hand, clips 15 and 16 were present in other highlights but not in V1. Single candidate quotes (clips 6,7,8) or back-and-forths between two candidates (clips 10,13,17) got picked up by more human highlights as well as our generic automated highlights video. On the other hand, issue-based partitions---such as keys 1,2,4,5,14,15,19--- the external highlights tended to agree less. The fact that the automated summary V1 covers clips from both these categories emphasizes the opportunity for personalization towards user taste or context.
%\block{There are some interesting stories for the overall V1 vs. all and the comparison of the different sources versus each other.  If you want to spend time comparing the different sources, this may speak to point (A) in the intro .  If you want to spend time talking about consumption differences (e.g. CNN vs. others) you can talk to point (C) from the intro}

% \subsection{Wimbledon 2019 Women's Final}\label{sec:tennis}
\boldpara{Wimbledon 2019 Women's Final.}
\begin{table}[t]
\centering
\scalebox{.9}{
\begin{tabular}{cl|ccc}
\hline
\parbox[t]{.75cm}{\centering{\bf Key}} &
\parbox[t]{4cm}{{\bf Games/points (W-H)}} &
\parbox[t]{1cm}{{\bf ESPN}} &
\parbox[t]{1.1cm}{\centering{\bf V1}} &
\parbox[t]{1.1cm}{\centering{\bf V2}} \\\hline \hline
1 &    0-2 /15-0              & Y     & Y   & Y      \\\hline
2 &    0-2 / 15-15            & Y     &  &     \\\hline
3 &    0-2 / 30-40            & Y     &  &     \\\hline
4 &    0-3                    &    & Y   & Y      \\\hline
5 &    1-4                    & Y     &  & Y      \\\hline
6 &    1-4 / 15-0             & Y     & Y   & Y      \\\hline
7 &    1-4 / adv H            & Y     &  &     \\\hline
8 &    2-5 / 0-0              &    &  & Y      \\\hline
9 &    2-5 / 0-15             &    &  & Y      \\\hline
10 &    2-5 / 15-15           & Y     & Y   & Y      \\\hline
11 &    2-5 / 30-40           & Y     &  &     \\\hline
12 &    2-6 0-0               &    &  & Y      \\\hline
%     \end{tabular}
% }}
% \subfloat{
% \centering
% \scalebox{.8}{
%     \begin{tabular}{cl|ccc}
% \hline
% \parbox[t]{.75cm}{\centering{\bf Key}} &
% \parbox[t]{2.3cm}{{\bf Games/points \newline (W-H)}} &
% \parbox[t]{1cm}{{\bf ESPN}} &
% \parbox[t]{1.1cm}{\centering{\bf V1}} &
% \parbox[t]{1.1cm}{\centering{\bf V2}} \\\hline \hline
13 &    2-6 0-0 / 0-15        & Y     & Y   & Y      \\\hline
14 &    2-6 1-1               &    & Y   &     \\\hline
15 &    2-6 2-1               &    &  & Y      \\\hline
16 &    2-6 2-1 / 0-15        &    & Y   & Y      \\\hline
17 &    2-6 2-1 / 0-30        & Y     & Y   & Y      \\\hline
18 &    2-6 2-2               &    &  & Y      \\\hline
19 &    2-6 2-2 / 0-15        & Y     & Y   & Y      \\\hline
20 &    2-6 2-2 / 15-30       & Y     &  &     \\\hline
21 &    2-6 2-2 / 15-40       & Y     &  &     \\\hline
22 &    2-6 2-4 / adv H       & Y     &  &     \\\hline
23 &    2-6 2-5 / 0-40        & Y     & Y   & Y      \\\hline
\end{tabular}
}
    \caption{Point-level comparison across highlights videos for the 2019 Wimbledon Women's final between Serena Williams (W) and Simona Halep (H).}
    \label{tab:wimtable}
\end{table}
The uniqueness of this media is that it has very clear time delineations, and that there is a higher temporal dependence (e.g. showing an end-game highlight before an early game score can spoil the results).
To test our automated methods (V1 and V2) against a specific external domain expert, we consider the highlights video available in the YouTube channel of ESPN \cite{espn}. For the event-based approach of V2, we obtain point-level timestamps and information for the match from a publicly available repository \cite{tennispbp}. To implement V2, we use the externally obtained timestamps to partition the game at the point level ($n=93$). We assign a score to each point by taking the mean of normalized rewatch scores for each second within that point boundary. Following this, we choose the 15 points with the highest scores as seed segments, perform shot-level smoothing of the seed clips and concatenate them to obtain the highlights video. The method for comparison is the same as the previous case study, by determining the presence or absence of clips in each highlights and comparing across highlights. However, determining intersection is simpler, since we have point scores for each player in place of text descriptions.

% \boldpara{Results}
Our content-agnostic variant highlights this media with clips from ten points---seven of them common with the ESPN version, and nine common with V2. The ESPN version selected clips typically at multiple scoring points in a game, indicated by keys 1--4, 5--7 and 19--21. This shows a preference of the domain expert curator towards narrative-based highlights. Interestingly, even though V2 was completely automated, it showed similar patterns, as indicated by the keys 5--6, 8--10, 12--13, 15--17 and 18--19. Finally, all methods picked up the match-ending point (key 23).

%\block{Can highlight the domain expert and the differences in variants here}

% \subsection{Discussion}
\boldpara{Discussion.}
To summarize the above case studies, while the public versions of the highlights videos tend to focus more on building up a story by including groups of closely-situated or correlated segments, not all segments within these groups generate high viewer interest per our analysis.  This generates a satisfactory answer to our first research question for the feasibility of a general algorithm for highlight generation for both correlated and uncorrelated events. Also, both algorithm variants generate comprehensive highlight overlap with public versions, answering the second research question of automated and curator parity. However, as with most automated systems, we assert that a the best system would be a
% While there are advantages to both approaches, it is possible to 
combined human-in-the-loop process by supplying our generated seed segments generated as inputs for a human curator while reducing the burden of domain expertise  across diverse content sources.

%% file: 4_conc.tex
\section{Conclusion and future work}\label{sec:conc}
We present an automated method for the generation of content highlights based on content analysis and viewership metrics.  Besides the repeat viewership metric, there are opportunities to use other metrics, for example quantifying or comparing between viewership feedback of multiple advertisements and correlating with their content characteristics using a timeline of tune-out percentages, or comparison of rewatch activities across multiple user segments. 

In the realm of content personalization, which is already challenging to accommodate in a curator environment, there is ever-growing in diversity of contexts for content consumption.  Although its growth was slowed by events in 2020, watching from mobile devices, during commutes, and varying friend and family social gatherings, has created new viewing contexts instead of classical theatrical and in-home environments \cite{nielsen_younger_2018}. We are interested in identifying more applications and forming new collaborations that can leverage this viewership-centric methodology. %A visualization portal for inspection of viewership metrics on 50 TV shows is available at blah. Curated (and in some cases personalized) highlights based on the high-interest segments is available at blah.

\begin{figure}[h]
    \includegraphics[width=.9\linewidth, clip, trim=70pt 70pt 70pt 70pt]{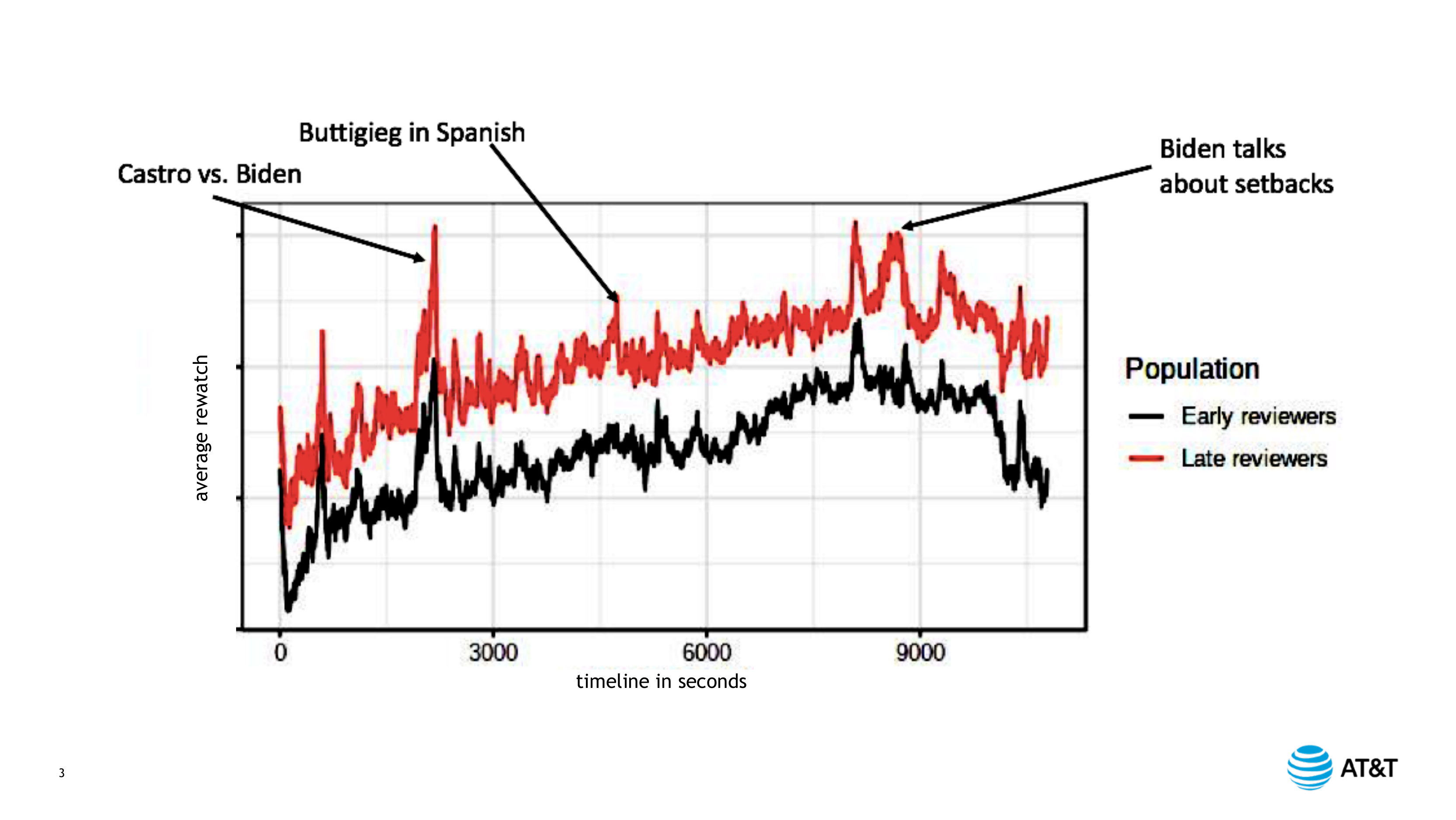}
    \caption{Behavioral difference in content rewatch patterns in viewership timelines of early and late reviewers of the Feb 6, 2020 Democratic Debate.}
    \label{fig:early-late}
\end{figure}

Finally, the technology outlined in this work presents new opportunities to generate a continuously evolving content %highly
stream. A preliminary experiment (Fig. \ref{fig:early-late}) surfaces interesting differences between the viewership timelines of early viewers (rewatch within 12 hours) vs. late viewers (rewatch between 1 and 2 days).  This provides a unique option for a dynamic curation process: generating an initial highlights video based on early viewing patterns, then updating it later to synchronize with possible media and social influence on late viewing behavior.

%% file: 0_template_sigchi.bbl

\begin{thebibliography}{25}

%%% the Web address.

\ifx \showCODEN    \undefined \def \showCODEN     #1{\unskip}     \fi
\ifx \showDOI      \undefined \def \showDOI       #1{#1}\fi
\ifx \showISBNx    \undefined \def \showISBNx     #1{\unskip}     \fi
\ifx \showISBNxiii \undefined \def \showISBNxiii  #1{\unskip}     \fi
\ifx \showISSN     \undefined \def \showISSN      #1{\unskip}     \fi
\ifx \showLCCN     \undefined \def \showLCCN      #1{\unskip}     \fi
\ifx \shownote     \undefined \def \shownote      #1{#1}          \fi
\ifx \showarticletitle \undefined \def \showarticletitle #1{#1}   \fi
\ifx \showURL      \undefined \def \showURL       {\relax}        \fi
% invisible to TeX
\providecommand\bibfield[2]{#2}
\providecommand\bibinfo[2]{#2}
\providecommand\natexlab[1]{#1}
\providecommand\showeprint[2][]{arXiv:#2}

\bibitem[\protect\citeauthoryear{??}{tim}{[n.d.]}]%
        {time}
 \bibinfo{year}{[n.d.]}\natexlab{}.
\newblock \bibinfo{booktitle}{\emph{{7 Democrats Face Off In The Last
  Democratic Presidential Primary Debate of 2019: Highlights}}}.
\newblock
\urldef\tempurl%
\url{https://time.com/5753060/debate-live-updates}
\showURL{%
\tempurl}


\bibitem[\protect\citeauthoryear{??}{cnn}{[n.d.]}]%
        {cnn}
 \bibinfo{year}{[n.d.]}\natexlab{}.
\newblock \bibinfo{booktitle}{\emph{{8 takeaways from the sixth Democratic
  presidential debate }}}.
\newblock
\urldef\tempurl%
\url{https://www.cnn.com/2019/12/20/politics/pbs-politico-debate-highlights/index.html}
\showURL{%
\tempurl}


\bibitem[\protect\citeauthoryear{??}{gua}{[n.d.]}]%
        {guardian}
 \bibinfo{year}{[n.d.]}\natexlab{}.
\newblock \bibinfo{booktitle}{\emph{{Democratic debate: sparks fly in the last
  debate of 2019 – video highlights}}}.
\newblock
\urldef\tempurl%
\url{https://www.theguardian.com/global/video/2019/dec/20/democratic-debate-sparks-fly-in-the-last-debate-of-2019-video-highlights}
\showURL{%
\tempurl}


\bibitem[\protect\citeauthoryear{??}{sla}{[n.d.]}]%
        {slam}
 \bibinfo{year}{[n.d.]}\natexlab{}.
\newblock \bibinfo{booktitle}{\emph{{IBM SlamTracker}}}.
\newblock
\urldef\tempurl%
\url{https://www.usopen.org/en_US/slamtracker/index.html}
\showURL{%
\tempurl}


\bibitem[\protect\citeauthoryear{??}{nym}{[n.d.]}]%
        {nymag}
 \bibinfo{year}{[n.d.]}\natexlab{}.
\newblock \bibinfo{booktitle}{\emph{{Key Moments From the December Democratic
  Debate}}}.
\newblock
\urldef\tempurl%
\url{http://nymag.com/intelligencer/2019/12/10-highlights-from-the-december-democratic-debate-wine-cave.html}
\showURL{%
\tempurl}


\bibitem[\protect\citeauthoryear{??}{nfl}{[n.d.]}]%
        {nflsavant}
 \bibinfo{year}{[n.d.]}\natexlab{}.
\newblock \bibinfo{booktitle}{\emph{{NFLSavant.com}}}.
\newblock
\urldef\tempurl%
\url{http://nflsavant.com/about.php}
\showURL{%
\tempurl}


\bibitem[\protect\citeauthoryear{??}{esp}{[n.d.]}]%
        {espn}
 \bibinfo{year}{[n.d.]}\natexlab{}.
\newblock \bibinfo{booktitle}{\emph{{Simona Halep vs Serena Williams Wimbledon
  2019 final highlights}}}.
\newblock
\urldef\tempurl%
\url{https://www.youtube.com/watch?v=VRzVd1OZoaQ}
\showURL{%
\tempurl}


\bibitem[\protect\citeauthoryear{??}{azu}{[n.d.]}]%
        {azure}
 \bibinfo{year}{[n.d.]}\natexlab{}.
\newblock \bibinfo{booktitle}{\emph{Video Indexer: Automatically extract
  advanced metadata from video and audio content}}.
\newblock
\urldef\tempurl%
\url{https://azure.microsoft.com/en-us/services/media-services/video-indexer/}
\showURL{%
\tempurl}


\bibitem[\protect\citeauthoryear{??}{nyt}{[n.d.]}]%
        {nyt}
 \bibinfo{year}{[n.d.]}\natexlab{}.
\newblock \bibinfo{booktitle}{\emph{{Watch: Highlights From the Democratic
  Debate}}}.
\newblock
\urldef\tempurl%
\url{https://www.nytimes.com/video/us/elections/100000006885259/democratic-debate-highlights.html}
\showURL{%
\tempurl}


\bibitem[\protect\citeauthoryear{{Amazon Rekognition}}{{Amazon
  Rekognition}}{[n.d.]}]%
        {aws}
\bibfield{author}{\bibinfo{person}{{Amazon Rekognition}}.}
  \bibinfo{year}{[n.d.]}\natexlab{}.
\newblock \bibinfo{booktitle}{\emph{Automate your image and video analysis with
  machine learning}}.
\newblock
\urldef\tempurl%
\url{https://aws.amazon.com/rekognition}
\showURL{%
\tempurl}


\bibitem[\protect\citeauthoryear{Baraldi, Grana, and Cucchiara}{Baraldi
  et~al\mbox{.}}{2016}]%
        {baraldi_recognizing_2016}
\bibfield{author}{\bibinfo{person}{Lorenzo Baraldi},
  \bibinfo{person}{Costantino Grana}, {and} \bibinfo{person}{Rita Cucchiara}.}
  \bibinfo{year}{2016}\natexlab{}.
\newblock \showarticletitle{Recognizing and {Presenting} the {Storytelling}
  {Video} {Structure} with {Deep} {Multimodal} {Networks}}.
\newblock \bibinfo{journal}{\emph{arXiv:1610.01376 [cs]}} (\bibinfo{date}{Nov.}
  \bibinfo{year}{2016}).
\newblock
\urldef\tempurl%
\url{http://arxiv.org/abs/1610.01376}
\showURL{%
\tempurl}
\newblock
\shownote{arXiv: 1610.01376.}


\bibitem[\protect\citeauthoryear{Basan}{Basan}{2020}]%
        {gotsentiment}
\bibfield{author}{\bibinfo{person}{E. Basan}.} \bibinfo{year}{2020}\natexlab{}.
\newblock \bibinfo{booktitle}{\emph{{Game of Thrones Character Sentiment
  Analyzer}}}.
\newblock
\urldef\tempurl%
\url{https://github.com/edfilbasan/gotsentiment}
\showURL{%
\tempurl}


\bibitem[\protect\citeauthoryear{Hodge}{Hodge}{2011}]%
        {anombook}
\bibfield{author}{\bibinfo{person}{V. Hodge}.} \bibinfo{year}{2011}\natexlab{}.
\newblock \bibinfo{booktitle}{\emph{{Outlier and Anomaly Detection: A Survey of
  Outlier and Anomaly Detection Methods}}}.
\newblock \bibinfo{publisher}{LAP LAMBERT}.
\newblock


\bibitem[\protect\citeauthoryear{Merler, Mac, Joshi, Nguyen, Hammer, Kent,
  Xiong, Do, Smith, and Feris}{Merler et~al\mbox{.}}{2019}]%
        {merler_automatic_2019}
\bibfield{author}{\bibinfo{person}{M. Merler}, \bibinfo{person}{K.~C. Mac},
  \bibinfo{person}{D. Joshi}, \bibinfo{person}{Q. Nguyen}, \bibinfo{person}{S.
  Hammer}, \bibinfo{person}{J. Kent}, \bibinfo{person}{J. Xiong},
  \bibinfo{person}{M.~N. Do}, \bibinfo{person}{J.~R. Smith}, {and}
  \bibinfo{person}{R.~S. Feris}.} \bibinfo{year}{2019}\natexlab{}.
\newblock \showarticletitle{Automatic {Curation} of {Sports} {Highlights}
  {Using} {Multimodal} {Excitement} {Features}}.
\newblock \bibinfo{journal}{\emph{IEEE Transactions on Multimedia}}
  \bibinfo{volume}{21}, \bibinfo{number}{5} (\bibinfo{date}{May}
  \bibinfo{year}{2019}), \bibinfo{pages}{1147--1160}.
\newblock
\showISSN{1941-0077}
\urldef\tempurl%
\url{https://doi.org/10.1109/TMM.2018.2876046}
\showDOI{\tempurl}


\bibitem[\protect\citeauthoryear{Nielsen}{Nielsen}{2018}]%
        {nielsen_younger_2018}
\bibfield{author}{\bibinfo{person}{Nielsen}.} \bibinfo{year}{2018}\natexlab{}.
\newblock \bibinfo{title}{Younger, {More} {Affluent} {Viewers} {Power}
  {Out}-of-{Home} {TV} {Viewing} {Lift}}.
\newblock
\newblock
\urldef\tempurl%
\url{https://www.nielsen.com/us/en/insights/article/2018/younger-more-affluent-viewers-power-out-of-home-tv-viewing-lift}
\showURL{%
\tempurl}


\bibitem[\protect\citeauthoryear{Otani, Nakashima, Rahtu, and Heikkilä}{Otani
  et~al\mbox{.}}{2019}]%
        {otani_rethinking_2019}
\bibfield{author}{\bibinfo{person}{Mayu Otani}, \bibinfo{person}{Yuta
  Nakashima}, \bibinfo{person}{Esa Rahtu}, {and} \bibinfo{person}{Janne
  Heikkilä}.} \bibinfo{year}{2019}\natexlab{}.
\newblock \showarticletitle{Rethinking the {Evaluation} of {Video}
  {Summaries}}.
\newblock \bibinfo{journal}{\emph{arXiv:1903.11328 [cs]}}
  (\bibinfo{date}{April} \bibinfo{year}{2019}).
\newblock
\urldef\tempurl%
\url{http://arxiv.org/abs/1903.11328}
\showURL{%
\tempurl}
\newblock
\shownote{arXiv: 1903.11328.}


\bibitem[\protect\citeauthoryear{Park, Rohrbach, Darrell, and Rohrbach}{Park
  et~al\mbox{.}}{2019}]%
        {park_adversarial_2019}
\bibfield{author}{\bibinfo{person}{Jae~Sung Park}, \bibinfo{person}{Marcus
  Rohrbach}, \bibinfo{person}{Trevor Darrell}, {and} \bibinfo{person}{Anna
  Rohrbach}.} \bibinfo{year}{2019}\natexlab{}.
\newblock \showarticletitle{Adversarial {Inference} for {Multi}-{Sentence}
  {Video} {Description}}.
\newblock \bibinfo{journal}{\emph{arXiv:1812.05634 [cs]}}
  (\bibinfo{date}{April} \bibinfo{year}{2019}).
\newblock
\urldef\tempurl%
\url{http://arxiv.org/abs/1812.05634}
\showURL{%
\tempurl}
\newblock
\shownote{arXiv: 1812.05634.}


\bibitem[\protect\citeauthoryear{Redacted}{Redacted}{2021}]%
        {snackable}
\bibfield{author}{\bibinfo{person}{Redacted}.} \bibinfo{year}{2021}\natexlab{}.
\newblock \showarticletitle{{Author names and title omitted due to the blind
  review}}.
\newblock


\bibitem[\protect\citeauthoryear{Retrosheet}{Retrosheet}{[n.d.]}]%
        {retro}
\bibfield{author}{\bibinfo{person}{Retrosheet}.}
  \bibinfo{year}{[n.d.]}\natexlab{}.
\newblock \bibinfo{booktitle}{\emph{Play-by-Play Data Files (Event Files)}}.
\newblock
\urldef\tempurl%
\url{https://www.retrosheet.org/game.htm}
\showURL{%
\tempurl}


\bibitem[\protect\citeauthoryear{Sackmann}{Sackmann}{2020}]%
        {tennispbp}
\bibfield{author}{\bibinfo{person}{Jeff Sackmann}.}
  \bibinfo{year}{2020}\natexlab{}.
\newblock \bibinfo{booktitle}{\emph{{Grand Slam Point-by-Point Data,
  2011-20}}}.
\newblock
\urldef\tempurl%
\url{https://github.com/JeffSackmann/tennis_slam_pointbypoint}
\showURL{%
\tempurl}


\bibitem[\protect\citeauthoryear{Smith}{Smith}{2016}]%
        {smith_ibm_2016}
\bibfield{author}{\bibinfo{person}{Smith}.} \bibinfo{year}{2016}\natexlab{}.
\newblock \bibinfo{title}{{IBM} {Research} {Takes} {Watson} to {Hollywood} with
  the {First} "{Cognitive} {Movie} {Trailer}"}.
\newblock
\newblock
\urldef\tempurl%
\url{https://www.ibm.com/blogs/think/2016/08/cognitive-movie-trailer/}
\showURL{%
\tempurl}


\bibitem[\protect\citeauthoryear{Vision}{Vision}{[n.d.]}]%
        {goog}
\bibfield{author}{\bibinfo{person}{Google Vision}.}
  \bibinfo{year}{[n.d.]}\natexlab{}.
\newblock \bibinfo{booktitle}{\emph{Vision AI}}.
\newblock
\urldef\tempurl%
\url{https://cloud.google.com/vision}
\showURL{%
\tempurl}


\bibitem[\protect\citeauthoryear{Wang, Liu, Puri, and Metaxas}{Wang
  et~al\mbox{.}}{2020}]%
        {wang_learning_2020}
\bibfield{author}{\bibinfo{person}{Lezi Wang}, \bibinfo{person}{Dong Liu},
  \bibinfo{person}{Rohit Puri}, {and} \bibinfo{person}{Dimitris~N. Metaxas}.}
  \bibinfo{year}{2020}\natexlab{}.
\newblock \showarticletitle{Learning {Trailer} {Moments} in {Full}-{Length}
  {Movies}}.
\newblock \bibinfo{journal}{\emph{arXiv:2008.08502 [cs]}} (\bibinfo{date}{Aug.}
  \bibinfo{year}{2020}).
\newblock
\urldef\tempurl%
\url{http://arxiv.org/abs/2008.08502}
\showURL{%
\tempurl}
\newblock
\shownote{arXiv: 2008.08502.}


\bibitem[\protect\citeauthoryear{Wongsuphasawat}{Wongsuphasawat}{2016}]%
        {wongsuphasawat_how_2016}
\bibfield{author}{\bibinfo{person}{Krist Wongsuphasawat}.}
  \bibinfo{year}{2016}\natexlab{}.
\newblock \bibinfo{title}{How every \#{GameOfThrones} episode has been
  discussed on {Twitter}}.
\newblock
\newblock
\urldef\tempurl%
\url{https://blog.twitter.com/en_us/topics/insights/2016/game-of-thrones-viz.html}
\showURL{%
\tempurl}


\bibitem[\protect\citeauthoryear{{Yale Song}, Vallmitjana, Stent, and
  Jaimes}{{Yale Song} et~al\mbox{.}}{2015}]%
        {yale_song_tvsum_2015}
\bibfield{author}{\bibinfo{person}{{Yale Song}}, \bibinfo{person}{Jordi
  Vallmitjana}, \bibinfo{person}{Amanda Stent}, {and}
  \bibinfo{person}{Alejandro Jaimes}.} \bibinfo{year}{2015}\natexlab{}.
\newblock \showarticletitle{{TVSum}: {Summarizing} web videos using titles}. In
  \bibinfo{booktitle}{\emph{2015 {IEEE} {Conference} on {Computer} {Vision} and
  {Pattern} {Recognition} ({CVPR})}}. \bibinfo{pages}{5179--5187}.
\newblock
\urldef\tempurl%
\url{https://doi.org/10.1109/CVPR.2015.7299154}
\showDOI{\tempurl}
\newblock
\shownote{ISSN: 1063-6919.}


\end{thebibliography}
